\documentclass[letter,preprint,aps]{revtex4}
\usepackage{graphics}
\usepackage{graphicx}
\usepackage{amsmath}
\usepackage{amssymb}
\begin{document}

%\begin{frontmatter}



\title{Theory Perspective: SCES '05 Vienna}
%---- Don't remove this comment line! ----

%\author[AA]{P. Coleman\corauthref{Name1}},
%\ead{coleman@physics.rutgers.edu}
\author{P. Coleman},

%\address[AA]{Center for Materials Theory,
%Rutgers University, Piscataway, NJ 08855, U.S.A. }
\address{Center for Materials Theory,
Rutgers University, Piscataway, NJ 08855, U.S.A. }

%\corauth[Name1]{Piers Coleman. Tel. 001 - 732 -445-5082
%fax: your number 1-732-445-4400}
\pacs {71.20.Lp, 71.27+a, 72.15.Qm,   74.70.Tx, 75.20.Hr, 75.30.Mb}
\begin{abstract}
Over the past  decades, research into 
strongly correlated electron materials 
that has consistently outperformed our wildest expectations, with 
new discoveries and radically new theoretical insights.  SCES '05
reaffirmed this vitality.  Highlight areas included 
parity violating superconductivity, a new, two dimensional  Helium-3
``Kondo lattice'' and the discovery of quantum
critical ferro-electricity. I review
these and many other developments, with an emphasis on open questions
and prospects for the future. 
\end{abstract}

%\begin{keyword}
%Summary\sep superconductivity \sep quantum phase \sep quantum
%criticality \sep heavy fermion
% PACS codes here, in the form: 
%\PACS 71.20.Lp, 71.27+a, 72.15.Qm,   74.70.Tx, 75.20.Hr, 75.30.Mb
%\end{keyword}
%\end{frontmatter}

% main text

%\eject
%
\maketitle
%
%\vfill\eject 

%\section{Introduction}\label{} 

% Extraordinary Science

% New Quantum Phases of Matter : URu2Si2;He-3; CePt_{3}Sn

% Theoretica

% Phenomenology

% New approaches ?

% Quantum Criticality

% The return of charge.

%--------------------------------------------------------
\section{Extraordinary Field}\label{}% 

Strongly Correlated Electron Systems, 2005 (SCES05) was held in
Vienna. This classical city, with its 
legendary coffee houses (many just outside the conference venue), 
proved a superb setting  to review and 
discuss the latest discoveries in this active
area of materials and condensed matter physics research. 

SCES05 is a direct descendant of the first International
Conference on Valence Fluctuations, held in Rochester New York in
1977\cite{rochester}.  
The 
consistency with which
the discovery of
new materials and new concepts 
has outstripped expectations is simply extraordinary. 
In 1977, the concept that localized f-electrons could form heavy
bands was treated with considerable skepticism
and the possibility of electronically mediated
superconductivity in dense local moment systems was regarded as highly
radical!
Our meeting in Vienna celebrated 30 years since these
seminal discoveries, yet it was also 
a forward looking conference, with a
host of incredible new discoveries. I came away from the meeting feeling like
a kid, with a huge sense of 
excitement about the prospects for the future. 

\begin{figure}[!ht]
\begin{center}
\includegraphics[width=0.65\textwidth]{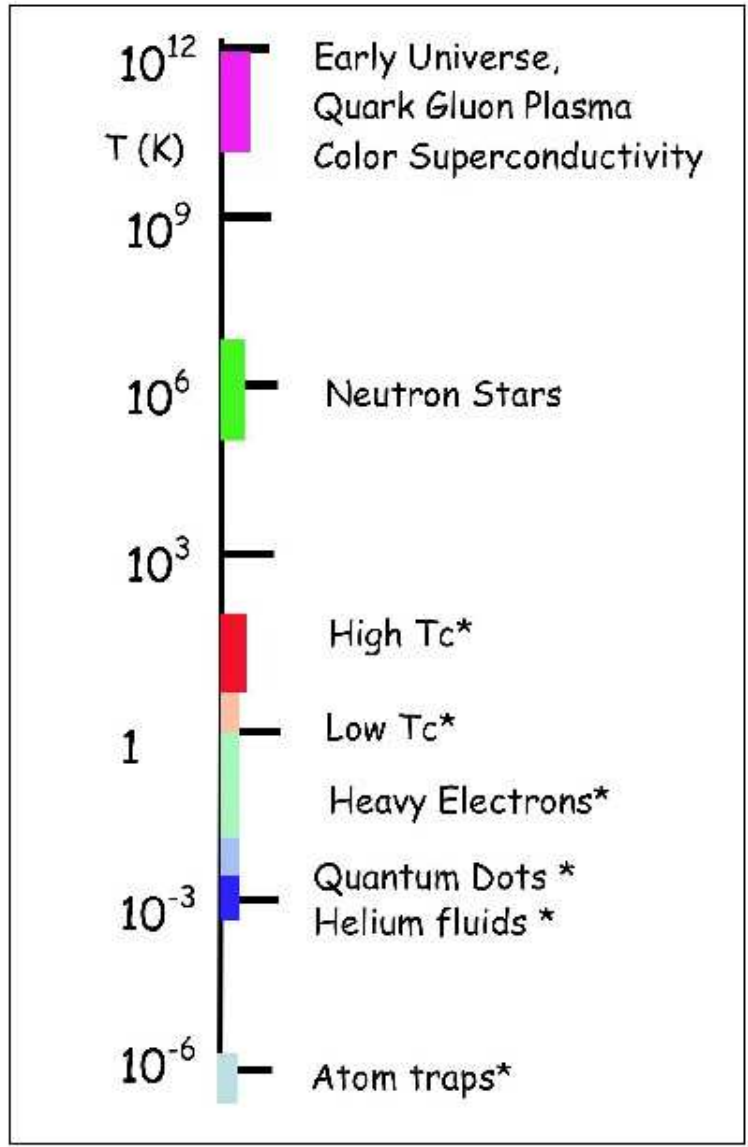}
\end{center}
\caption{Strongly correlated behavior of matter and its relationship to
the research presented at this meeting (starred). 
Although the range 
of energy where research is carried out spans from roughly $10^{4}K$
to $10^{-3}K$, ideas and research on strongly correlated electron
systems in the lab have
a broad connection with physics on far lower, and far greater energy
scales. }
\label{fig1}
\end{figure}

Strongly correlated electron physics
is a field
never far from applications, and 
strong links between research and
the development of rare earth magnets, thermo-cooling, cryogenics, multiferroics and
applied superconductivity continue today. 
We often forget however the equally important role of this 
field of physics as frontier science in the ``middle ground'' of our quantum universe.
The domain of experiments reported at this conference 
spans a temperature range from room temperature  to millikelvin,
and within this playground, we have the opportunity to explore the
general principles that govern collective behavior in matter, often using discoveries made
on one scale, to understand the physics on another.
For example, discoveries made in actinide and rare earth 
``heavy electron physics'' have had broad influence
on our understanding of magnetism and high temperature superconductivity in 
more complex oxide materials, yet they have also influenced research 
into quantum dots and mesoscopics. 
But  the discoveries and ideas of
strongly correlated electron physics also enjoy 
an influence that stretches far further along the energy axis, on one side
out to the  physics of the early universe,
neutron stars,  quark gluon plasmas and anisotropic color
superconductivity, and on the other side, 
down to the low energy physics of 
Helium-3 and the tiny energy scales of atom
traps (Fig. \ref{fig1}. ). 
Today, experimental discovery and 
theoretical ideas developed through strongly  correlated electron systems
continue to enjoy a strong mutual influence
with these kindred fields of research. 

\vskip 0.2cm

\section{New Quantum Phases of Matter : $URu_{2}Si_{2}$,  $CePt_{3}Sn$ and
bilayer $^{3}He$}

One of the themes of this field, is the quest to discover
new quantum phases of matter and realize them 
in practical materials. Along the way, we need to learn the guiding
principles that help us to navigate 
amongst the vast space of the periodic table.
There were many new discoveries of new phases and materials 
reported at this meeting, and I'd like to highlight three of them. 
\begin{itemize}
\item $URu_{2}Si_{2}$, where
a ``nexus'' of new phases appear to cluster around a field-induced
quantum critical point\cite{nexus,harrison,satovienna,bernal_vienna,zhitormsky}.

\item $CePt_{3}Si$ a broken parity superconductor,
containing a coherent admixture
of singlet and triplet Cooper pairs\cite{bauer,cept3si1,cept3si2,cept3si3}, 

\item Bilayer $^{3}He$ on graphite, which enters the scene as
a new class of two dimensional heavy fermion fluid\cite{saunders2,saunders1}. 

\end{itemize}

\begin{figure}[!ht]
\begin{center}
\includegraphics[width=0.65\textwidth]{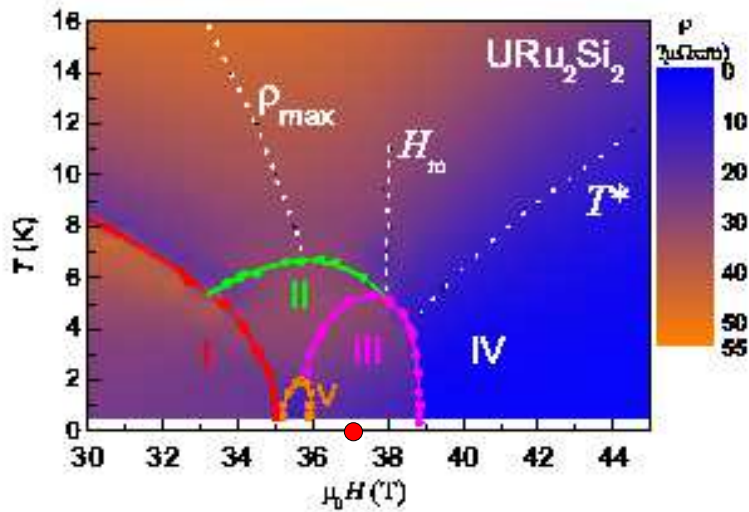}
\end{center}
\caption{Nexus of critical points in $URu_{2}Si_{2}$ determined from a
combination of magneto-caloric and resistivity measurements\cite{nexus}. 
Fig. courtesy of N. Harrison. Color coding is determined by the resistivity. 
The
extrapolation of $T^{*}$ to zero (red circle) is thought to be 
a quantum critical end point associated with a metamagnetic
transition. Phase (I) is the ``hidden order'' phase of
$URu_{2}Si_{2}$. Phase (IV) is a fully polarized Fermi liquid. 
 }
\label{fig2}
\end{figure}
A key observation of experimentalists, 
is that new phases of matter
tend to accumulate in the vicinity of quantum critical points. The 
heavy electron superconductor $URu_{2}Si_{2}$ provides a fascinating
realization of this phenomenon. 
Kim,
Harrison et al.\cite{harrison} report on a plethora of new phases that develop 
in $URu_{2}Si_{2}$ upon application of high magnetic fields (Fig. \ref{fig2}. ). 
They have discovered that the 
high field specific heat of this material (doped with rhodium to 
suppress some of the order around the critical point) closely resembles
that observed near a metamagnetic phase transition 
in $CeRu_{2}Si_{2}$. Metamagnetism involves
a sudden cross-over magnetization  from unpolarized to
fully polarized Fermi liquid.  Studies of the quasi two dimensional
material $Sr_{3}Ru_{2}O_{7}$ 
suggest that metamagnetism takes place in materials
that lie close to the critical end-point of a line of first order
``ferromagnetic'' phase transitions\cite{grigera}.  
A new analysis of Harrison et al.
provides further support for the idea
that quantum critical end points can act as nucleating point for
new quantum phases. 

I was also excited by new progress towards identifying the 
the ``hidden order''  (HO) that develops in this  $URu_{2}Si_{2}$
below 17K\cite{hidden_order}. It is this order which 
is persists 
up to the high fields where nexus of new phases begin. 
Three new pieces of insight about the HO were announced at 
Vienna. 
It is now increasingly clear that the HO state is distinct from a
large moment antiferromagnetic phase that phase-coexists
with the hidden order under pressure. At this meeting, 
Sato et al\cite{satovienna} showed measurements of the susceptibility
under pressure which indicate 
that heavy fermion superconductivity is unique to the hidden order
phase, and does not develop in the large moment antiferromagnet. 
Oscar Bernal et al. presented \cite{bernal_vienna}
NMR measurements on single crystal samples of this material
that confirm a growth of tiny local fields inside the
hidden order phase that had previously been seen in powder experiments\cite{bernal_old}.
Remarkably, these local fields -
about 4-8 Gauss, do not change in a 14.5 Tesla field. 
This field-insensitivity adds support to the idea that these tiny fields
derive from staggered orbital currents  rather than from spin dipoles.
In a separate development, 
Zhitormsky and Mineev\cite{zhitormsky} described how, in their recent work 
on the band-theory of $URu_{2}Si_{2}$ they have been able to newly identify
nested regions of the Fermi surface, which are strong candidates 
as regions for the development of the hidden order. 
Zhitormsky and Mineev propose a novel kind of spin order, 
with a highly renormalized $g-$ hidden order. But whether the hidden
order is spin, 
or orbital order, the identification
of this nested Fermi surface as the origin of the Fermi surface
gapping that takes place at $17K$ is an important step forward in this
twenty year old mystery. 

\begin{figure}[!ht]
\begin{center}
\includegraphics[width=0.65\textwidth]{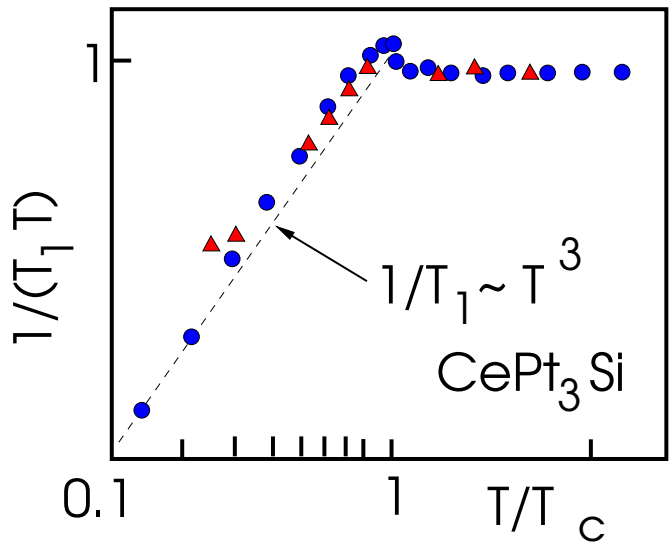}
\end{center}
\caption{Sketch of the temperature dependence of $1/T_{1}T$
(normalized with respect to its value at $T_{c}$) for 
the parity violating heavy electron superconductor, $CePt_{3}Si$ 
after \cite{cept3si2nmr}. }
\label{fig3}
\end{figure}

One of the new developments in superconductivity concerns
$CePt_{3}Si$\cite{bauer,cept3si1,cept3si2,cept3si3}.  
This new heavy fermion superconductor is
part of a growing trend of condensed matter interest in materials that
break parity inversion symmetry.  $CePt_{3}Si$\cite{bauer} has a
non-centrosymmetric crystal structure, so that unlike most
superconductors, the parity of the Cooper pairs is not a good quantum
number\cite{cept3si1}. The pair correlation function in a
superconductor can be written
\begin{equation}
F_{\alpha \beta } (x)
=\langle \psi _{\alpha}(x)\psi_{\beta}(0)\rangle  
\end{equation}
where 
\begin{equation}
F_{\alpha \beta } (x)=
\left[ i
\sigma_{2} ( F_{s} (x) +
\vec {\sigma} \cdot\vec{F}_{p} (x))
\right]_{\alpha \beta }
\end{equation}
is expanded in terms of a singlet, even-parity pair function
$F_{s} (x)= F_{s} (-x)$ and a triplet, odd-parity pair function
$\vec{F}_{p} (x) = - 
\vec{F}_{p} (-x)
$.
When parity is a good quantum number, 
either $F_{s}$ or $F_{p}$ are zero. 
In $CePt_{3}Si$, the absence of parity conservation implies
that both singlet and triplet pairs 
coherently coexist.  Experimental and
theoretical work on this material has progressed at an impressive rate
in the two years since its discovery. Mamoru Yogi presented new NMR
data\cite{cept3si2} which 
provides
a striking confirmation of this co-existence of triplet and singlet
pairs, exhibiting a small coherence peak characteristic of s-wave
superconductivity, slowly crossing over to a $T^{3}$ dependence 
at low
temperatures that is characteristic of line nodes around the Fermi
surface (Fig. \ref{fig3} ). Agterburg, Frigerie and collaborators\cite{cept3si1,cept3si2} have developed a 
phenomenological theory of this superconducting state which can
account for the NMR data. In their model, 
parity violation enters as a spin-dependent hopping.

The microscopic theory of heavy
electron superconductors, is a still a wide open question, and
$CePt_{3}Si$ joins a long list of bad-actor superconductors, including
$UBe_{13}$, $CeCoIn_{5}$ and $PuCoGa_{5}$ 
for which there is almost no microscopic understanding. 
We do know that the entropy associated with 
the Fermi liquid and the superconductor derives from f-spin entropy.
Heavy electron quasiparticles are really composite combinations of f-spins
and conduction electrons. The inclusion of this 
``Kondo physics '' into heavy electron superconductivity is a largely
unexplored theoretical challenge.  One of the paradoxes that needs to
be explored concerns the strong Coulomb repulsion between the f-electrons
which we expect to severely suppress the s-wave pairing component of the
f-electrons, so that 
\begin{equation}
\langle f^{\dagger }_{k\uparrow}f^{\dagger}_{-k\downarrow}\rangle \sim 0
\end{equation}
In a Kondo lattice, this suppression is total, because of the local
$SU (2)$ gauge symmetry associated with the f-spins\cite{affleck,colemanandrei}. 
From this perspective, it is difficult to understand how 
the pair-condensation entropy can be associated with s-wave pairing
and f-electrons at the same time, and this is an interesting paradox
to motivate future work. 

One of  the most surprising discoveries announced at SCES05,
is the discovery of heavy-fermion behavior in bilayer $^{3}He$ films
on graphite\cite{saunders2,saunders1}.
In the bulk, $^{3}He$ is {\sl the } historic paradigm for Fermi liquid behavior
and anisotropic superfluidity\cite{landau}. 
Helium films adsorbed on graphite
form two dimensional quantum fluids\cite{greywall}, in which the Helium atoms
move coherently across the 
periodic triangular  potential of the graphite surfaces. 
These are very clean, model strongly correlated fermi systems, 
with no complications from crystal fields or the
separate electronic and nuclear
contributions to the specific heat.

\begin{figure}[!ht]
\begin{center}
\includegraphics[width=0.65\textwidth]{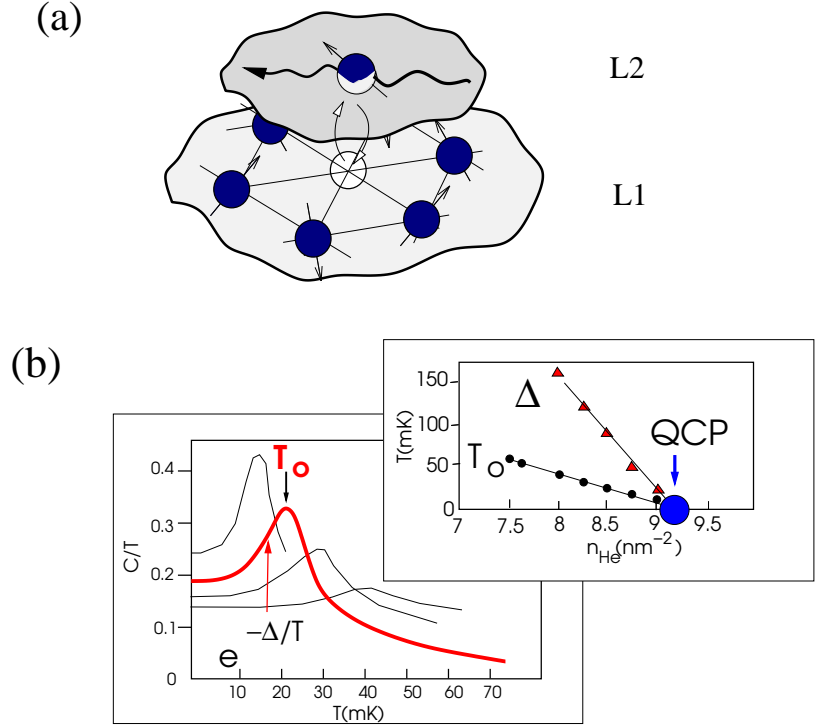}
\end{center}
\caption{(a) Schematic of bilayer $^{3}He$, showing almost localized
lower layer of spins and delocalized upper ``conduction sea''.
Valence fluctuations between the two layers melt the lower layer to produce a two dimensional
heavy fermion fluid. (b) Sketched evolution of  $C_{v}/T$ \cite{saunders1}
as a
function of filling, showing peak temperature $T_{0}$,  and
exponential decay $e^{-\frac{\Delta }{T}}$ that are used to define the
characteristic energy scales. 
Inset: collapse of characteristic scales to zero
at a quantum critical point where  $n= 9.2 \hbox{nm}^{-3}$. 
 }
\label{fig4}
\end{figure}

John Saunders \cite{saunders2} described how his group had succeeded in
measuring the properties of bilayer $^{3}He$ on graphite. 
In previous work\cite{saunders1}, they established that 
monolayer
$^{3}He$ films form a Fermi liquid which undergoes a Mott transition into
a triangular lattice solid at a critical filling factor $n_{He}\sim 5nm^{-2}$.   
Using a slightly modified substrate, the same
group has now succeeded in adding more $^{3}{He}$ atoms to form a bilayer
quantum fluid. The new bilayer system remains a Fermi liquid up to
a much higher critical coverage around  $9.9nm^{-2}$, where it undergoes
a quantum phase transition into a magnetic state.
Saunders notes that bilayer $^{3}He$ may closely resemble 
a two-dimensional heavy fermion system, in which the lower layer of
atoms is an almost localized spin system and the upper layer behaves
as a conduction band.Zero point ``valence fluctuations'' between the layers may play
an important role, causing spin exchange between the upper and lower
fluid that closely resembles the physics of a two dimensional Kondo
lattice( \ref{fig4}. (a)). 
The observed specific heat curves show a maximum
$T_{0}$ and appears to indicate the presence of an as-yet unexplained
gap $\Delta $ in the excitation spectrum.
As the filling of the
second layer is increased, the inverse mass of the Fermi liquid
appears to collapse linearly to zero as the critical coverage is
approached, while $T_{0}$ vanishes with a power law. Mysteriously, the
gap appears to drop to zero at a lower value of the filling
around $n_{c}\sim 9.2nm^{-2}$(Fig. 4 (b)). 
This is an exciting new discovery that should be a great
encouragement to the field, prompting new theoretical work and tending
to support the idea that the physics of quantum criticality seen in
electronic heavy fermion systems may in fact, be universal to a much
broader class of fermionic matter.

%%  Vienna Summary talk outline.
%%  
%%  1.  Overview of talk
%%  
%%  Extraordinary Field
%%  New Quantum States of Matter

\section{Theoretica!}\label{}

One of the dreams of numerical
approaches, is to develop tools that on the one hand, can provide us
with new insights into simple models but on the other hand, can 
guide us in the discovery of materials with
new quantum properties. 
There is a strong tradition of creative
numerical algorithms in the field of strongly correlated electrons,
one that stretches back to Ken Wilson's pioneering numerical
renormalization group solution to the Kondo model\cite{wilson}.  
This tradition continues today.

A modern descendant of Wilson's numerical
renormalization group approach is the density matrix renormalization
group (DMRG)\cite{white}. For one-dimensional models, this 
has become a state-of-the-art tool. 
An example of beautiful work in this direction, is provided by the
DMRG study of the quarter filled, 
one dimensional Kondo lattice, by Hotta and Shibata
\cite{hotta}.  Earlier DMRG work on this topic had proposed that at a quarter
filling, the Kondo lattice becomes unstable to an insulating state,
with with long-range staggered dimer order  
\begin{equation}
\langle \vec{S}_{j}\cdot
\vec{S}_{j+1}\rangle \sim \Psi e^{ i \pi j} .
\end{equation}
Hotta and Shimuzu have
carefully studied the size dependence of dimer correlations using DMRG,
triangulating their results by
comparing even and odd-numbered chains.  They conclude 
that the dimer order has power-law correlations, but no long range
order. This work illustrates how, despite great advances, very careful scaling
studies continue to be absolutely vital before drawing hard and fast
conclusions about the thermodynamic limit.

Continued progress is also being made in the application
of direct diagonalization approaches. 
One of the unsolved pieces of physics that challenges
the theoretical community is the Doniach quantum phase transition
that is believed to separate 
local moment magnetism, and heavy electron
paramagnetism in the metallic Kondo lattice\cite{doniach}.  
Zerec et al presented a
finite temperature ``Lanczos'' diagonalization study of the two
dimensional Kondo lattice\cite{zerec}, 
in which the temperature dependence of the
specific heat shows a two-peak structure that may signal the
approach to the magnetic quantum critical point. I have the sense that
these approaches will be amongst the first 
to provide us with the first detailed
phase diagram of the two dimensional Kondo lattice. 

One area of immense interest, is the application of dynamical mean
field theory (DMFT) to strongly correlated systems.  
Dynamical mean field theory\cite{dmft} regards
strongly correlated electron systems as an effective impurity, or
cluster of atoms, embedded in a self-consistently determined
environment. This approach is in many ways, the 21st century
descendant of density-functional theory, but now, the Free energy is
now a functional of the local, or cluster approximated Green's functions. 
The challenge
becomes to efficiently 
compute the functional of the local or cluster Green's functions.

Many interesting papers were presented which marry the dynamical mean
field theory with local, or cluster ``solvers'' provided by
a variety of other methods. 
For example, Koller and Hewson\cite{koller} showed how the  combination of DMFT, with the
use of numerical renormalization group to provide the impurity solver,
could provide new insight into
strong polaronic renormalization of electron masses in a
Holstein-Hubbard model. 

In an innovative piece of work,
Hanke, Aichorn, Dahnken, Arrigoni and Potthoff\cite{hanketal} demonstrated how
they can extract the functional dependence of the free energy on
cluster Greens functions from a Monte Carlo study of clusters. The
function is
then used to compute the translationally invariant lattice Green's functions. 

Dieter Vollhardt\cite{vollhardt} showed how the older density
functional theory can be combined with the DMFT, and 
illustrated the tremendous potential of
these hybrid methods for the study of metal-insulator transitions in vanadium
oxide systems, notably $SrVO_{3}$.  Vollhardt showed how these methods
were able to track both the incoherent local background, and the dispersive
component of the electron spectral functions, which contained ``kinks''
in the dispersion, corresponding to mass-renormalization effects. 
Frontier research of this sort is, I believe, slowly providing the 
framework for ab-initio many body studies of the future.

\section{New Phenomenology}

A vital middle ground for research in our field, is
the development of new, phenomenological approaches. Condensed matter
physics is driven largely by experimental discovery. 
Phenomenology is not only a stepping stone to deeper theoretical
insight, it is also the key  to a  closer interplay between theory and
experiment.

Two interesting pieces of phenomenology caught my eye at SCES05. 
Achim Rosch\cite{rosch} discussed ``Gr\" uneisen'' parameter $\Gamma_{T}$
can be used to characterize pressure or field-tuned
quantum phase transitions. 
The Gr\" uneisen parameter is the ratio of the thermal
expansivity $\beta = V^{-1}\partial V/\partial T$ to the specific, heat $C_{P}$
\begin{equation}
\Gamma_{T}= \frac{\beta }{C_{P}}.
\end{equation}
Near a quantum critical point, the characteristic scale of the Fermi
liquid $T_{0} (x)\sim x^{\nu z}$. collapses to zero as power of the separation $x$
from the critical point. Simple scaling analyses show that
the collapse of this scale to zero leads to a 
a divergence of the temperature dependent 
Gr\" uneisen parameter at a quantum critical point, given by
$\Gamma_{T}\sim 1/ T^{1/\nu z}$\cite{roschsi}. 
Rosch showed how degree of divergence of 
the Gruneissen parameter  is an important method to delineate between
``local'' and ``spin density wave'' magnetic quantum  critical
points.  
Moreover, the Gr\" uneissen parameter should 
should change sign at a quantum critical
point, making it an extremely useful tool for hunting for quantum phase transitions.

One of the most useful unifying trends in strongly correlated electron
systems is the Kadowaki Woods relation\cite{kadowakiwoods}
\begin{equation}
\frac{A}{\gamma^{2}}= W\sim 1 \times  10^{-5}{\mu \Omega cm
}\left(\frac{K mol}{mJ} \right)^{2}
\end{equation}
between the quadratic ``$A$''
coefficient of the resistivity ($\rho  = \rho_{0}+ AT^{2}$) and the
specific heat coefficient $\gamma= C_{V}/T$. 
The constancy of this quantity over a wide range of
strongly correlated materials is a reflection of the local character
of the quasi-particle interactions. 
Tsujii et al. \cite{tsujii} have made the
observation that several $Yb$ compounds with a large orbital
degeneracy, develop a consistently lower Kadowaki Woods ratio. 
In a lovely piece of phenomenology, in which they analyze the
dependence of the Fermi liquid properties on spin degeneracy $N$,
Tsujii et al find that both $A$ and $\gamma$ scale as $N (N-1)$\cite{tsujii2},
\begin{equation}\label{}
A  \propto\left(\frac{N (N-1)/2}{T_{0}^{2}} \right),
 \qquad \gamma \sim \left(\frac{N (N-1)/2}{T_{0}} \right),
\end{equation}
where $T_{0}$ is the characteristic scale of the Fermi liquid, 
so that the Kadowaki Woods relation becomes
\begin{equation}
\frac{A}{\gamma^{2}} \times \frac{N (N-1)}{2}= W\sim 1 \times  10^{-5}{\mu \Omega cm
}\left(\frac{K mol}{mJ} \right)^{2}
\end{equation}
This  degeneracy dependence 
accounts for the low KW ratio in $Yb$ compounds, and when taken into
account, brings them into line into align with the broad majority
of lower spin degeneracy materials.

There were so many fascinating discoveries I encountered in the
meeting that seem prime fodder for 
future phenomenological work. Here are 
three phenomena that may fall into this category:
\begin{itemize}

\item Gegenwart pointed out a fascinating property of the  Fermi
liquid ground-state of  $YbRh_{2}Si_{2}$, as it is field tuned towards
the quantum critical point\cite{gegenwart1,gegenwart2}. Although 
the specific heat coefficient $\gamma = C_{V}/T$,
$\chi $ and the quadratic coefficient of the resistivity all
diverge, it is the modified Kadowaki Woods ratio 
$A/\chi^{2}$ rather than $A/\gamma^{2}$ that remains constant.
What simple model for the momentum-dependent interactions in a Fermi liquid model can account
for this scaling? 

\item At the metal to Kondo-insulator phase transition observed by
Slabarski and Spalek\cite{slebarski1,slebarski2} in tin doped $CeRhSb$, the conductivity $\sigma $
scales with the magnetic susceptibility $\sigma/\chi = \hbox{cons}$. Why?

\item In the heavy electron superconductor $CeCoIn_{5}$, a
Fulde-Ferrell-Larkin-Ovchinikov (FFLO)
phase, (in which the  superconducting order parameter is spatially modulated)
is observed at low temperatures, near the Pauli-limited upper
critical field\cite{fflo1,fflo2,radovan,fflo3}.  As the field is rotated away from the c-axis, a
sequential heirachy of FFLO phases develop\cite{fflo1,radovan}, 
manifested by a sequence of plateau's in the magnetization.
This physics does not  fit into the existing understanding of FFLO
phases, and it may involve the stabilization of vortex structures 
through the development of antiferromagnetism
in the vortex cores.  Traditionally, FFLO phases
are treated using non-uniform solutions of the Boguilubov de Gennes
equation\cite{fflobdg}, but the interplay with magnetism makes such an approach far
too complicated for these systems. Would it be possible, I wonder, 
to develop a Landau Ginzburg theory for FFLO phases that can take
antiferromagnetic vortex cores into account? 

\end{itemize}

\section{Charge and Quantum Criticality.}\label{}

Quantum criticality was a 
very central topic at this meeting. 
Zachary Fisk has likened quantum critical points to a kind 
of black-hole - an essential singularity located at absolute zero in
the material phase diagram\cite{qc1,qc2}. 
We're fascinated by quantum criticality 
in part because
\begin{itemize}

\item as we saw in the case of $URu_{2}Si_{2}$, 
materials quantum critical points provide a kind of nucleation point
for the formation of new phases of quantum matter\cite{nexus,grigera}. 

\item 
quantum criticality
influences a broad swath of the material phase diagram, leading, to
the formation of metals with very unusual transport properties -
variously refereed to as ``strange'', or ``non- Fermi liquid'' metals\cite{review}.

\item of the possibility of new universality classes of critical
behavior that lie outside those previously observed at classical
critical points. 

\end{itemize}

SCES '05 introduced a number of fascinating new factors into quantum
criticality.  There were 
two new examples of quantum criticality that 
involve charge, rather than spin order:
\begin{enumerate}

\item Quantum critical ferro-electricity\cite{rowley}.

\item Critical valence fluctuations and their possible relation with
unconventional superconductivity.

\end{enumerate}

\begin{figure}[!ht]
\begin{center}
\includegraphics[width=0.65\textwidth]{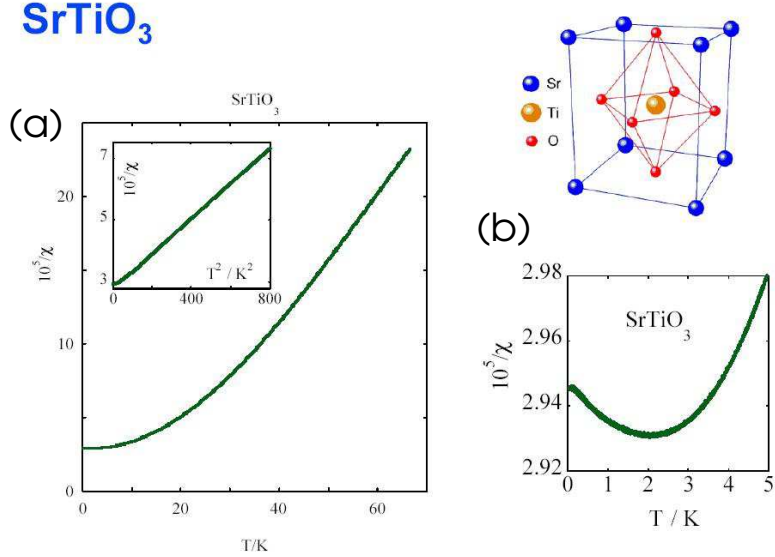}
\end{center}
\caption{Temperature dependence of the dielectric constant of $SrTiO_{3}$
after Stephen Rowley et al.\cite{rowley}, showing (a) cross-over from
$1/2$ Curie behavior at high temperature to $1/T^{2}$ divergence
at low temperature (b) small low temperature maximum that  believed
to be associated with a first order transition. 
 }
\label{fig5}
\end{figure}

In a post-deadline poster, Stephen Rowley et al\cite{rowley} described the
measurement of a divergence of the dielectric constant in strontium
titanate that follows a $1/T^{2}$ behavior, rising up to values in
excess of $\chi = 30,000$ in the approach to a ferro-electric quantum
critical point (Fig.\ref{fig5}). Strontium titanate may offer the
possibility of a text-book quantum phase transition, because
there are no complicating effects of electron damping in insulators.
In this system,  the temporal quantum fluctuations of the electric polarization
should count as one additional dimension, 
so the 
quantum critical behavior in strontium titanate should be equivalent to an Ising phase
transition in four-dimensions. This is the marginal, or ``upper critical'' dimension
for the Ising model. 
Rowley et al point out that the
$1/T^{2}$ divergence may be understood in terms of Gaussian, quantum
fluctuations in the dipole moments that exist  beyond four dimensions, 
but that there may be important marginal
logarithmic corrections associated with the critical dimensionality.

\begin{figure}[!ht]
\begin{center}
\includegraphics[width=0.65\textwidth]{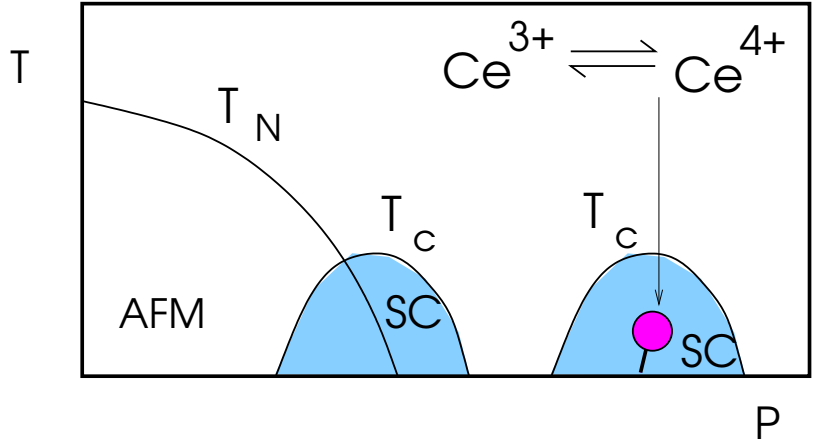}
\end{center}
\caption{Schematic phase diagram for $CeCu_{2} (Ge,Si)_{2}$ after
\cite{miyake}
illustrating possible valence fluctuation critical point beneath the
superconducting dome at high pressures. 
 }
\label{fig6}
\end{figure}

Kazumasa Miyake\cite{miyake} advanced a fascinating 
proposal that the superconducting phase diagram (Fig. \ref{fig6} ) of
$CeCu_{2} (Ge,Si)_{2}$ exhibits a maximum in the transition
temperature in close vicinity to a valence-changing critical
point\cite{cecu2gesi}. 
Finite temperature critical end-points associated with first
order valence changing phase transitions, such as the $\alpha -\gamma$
transition in elemental cerium, been known for decades. Miyake argues
that superconductivity may develop around the
region where the critical end point is suppressed to zero, to become a
quantum critical point. 

To these  new developments, I should also add the recent discovery of a
charge Kondo effect in Thallium doped $PbTe$
by a group at Stanford\cite{fisher,dzero}, where charge $2e$ fluctuations of the thallium ions
give rise to a charge scattering analog of the Kondo effect. This is
the charge $2e$ companion to the phenomena discussed by Miyake. 
These new developments make me believe that we have underestimated
the importance of slow charge fluctuations at quantum critical points,
and that charge will play an increasingly important role in our
consideration of quantum criticality in the years ahead.

There are substantial challenges
that still face us, in trying to understand magnetic quantum critical
points.  The idea of a phase transition between the heavy electron
paramagnet and the local moment antiferromagnet dates back to
Doniach's original proposal\cite{doniach} at the Rochester conference, that heavy electrons are dense Kondo
lattice systems.  Despite the vintage of this idea, we still do not have
a theory for how 
zero-point spin fluctuations melt antiferromagnetic order in a
metal. Indeed, our understanding of such quantum phase transitions is
far better evolved for insulating antiferromagnetic order, than it is
for metallic antiferromagnets.  At this meeting, Qimiao Si\cite{qimiao} emphasized
how many of the properties of quantum critical heavy electron systems
can not be understood in terms of a Hertz and Moriya spin fluctuation
picture of quantum phase transitions\cite{hertz,moriya}. 
Si, Ingersent, Smith and Rabello\cite{si}
have proposed that the bad-actor heavy electron quantum critical
points are quasi-two dimensional spin fluids in which the important
critical spin fluctuations are critically correlated in time.
Local quantum criticality is certainly the most mature theory of heavy
electron quantum criticality that we have at the current time. Many
other ideas have also been advanced as contenders. One of the ideas
floating around the community, is that the composite electrons
which form the heavy electron fluid may break into spinons  and holons
at the quantum
critical point. This has been an idea explored by Catherine Pepin in
her recent work\cite{pepin05}.  The idea of deconfined spinons also features in the
ideas of ``deconfined criticality'' advanced by Senthil et al.\cite{senthilfisher}.
None of these approaches is yet able to provide
a clear idea of how to connect the a state of three dimensional magnetic order to the 
three-dimensional heavy electron paramagnet.

My own belief is that progress on quantum criticality may benefit
by partially refocusing our effort towards developing a mean-field theory
to connect local moment magnetism with the heavy fermion metal.  Just
as Landau and Weiss mean theory of ferromagnetism provided the key
insights into the soft modes of classical criticality, I suspect that
a proper mean-field theory linking the Kondo lattice to local moment
magnetism will provide key insight into the unusual zero modes
responsible for the quasi-linear resistivity $\rho \sim T^{1+\eta }$
and the 
unusual logarithmic
temperature dependence of the specific heat coefficient $C_{v}/T\sim
(\frac{1}{T_{0}})\ln (T_{0}/T)$ seen in heavy electron
criticality\cite{review}.

\begin{figure}[!ht]

\begin{center}
\includegraphics[width=0.65\textwidth]{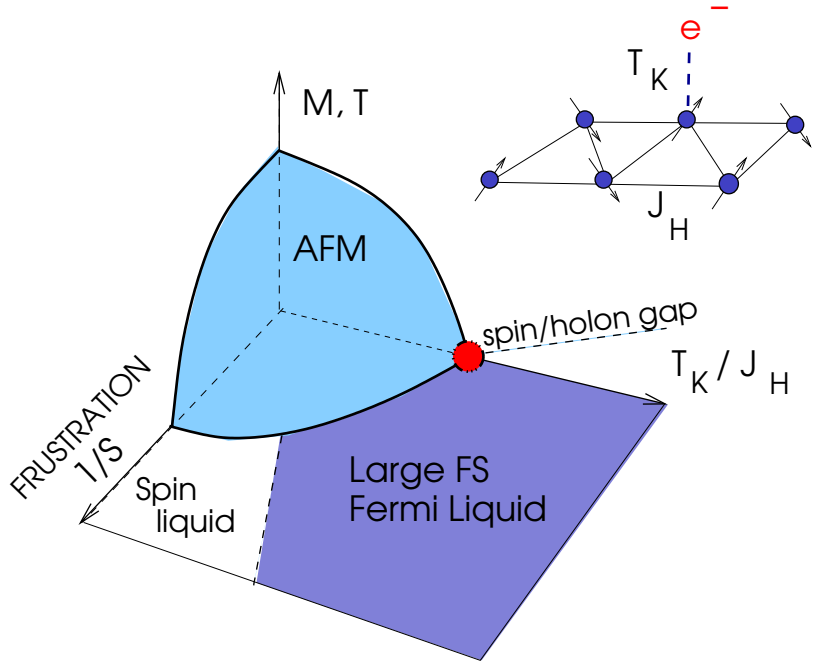}
\end{center}
\caption{Connection between the phase diagram of insulating spins with
the Kondo lattice. 
 }
\label{fig7}
\end{figure}

The discovery of heavy fermion behavior in bilayer $^{3}He$ announced
at this conference\cite{saunders2} provides tremendous new impetus to theoretical studies
of heavy fermion criticality. Here we have an absolutely two dimensional
heavy fermion system to which we may compare our theories.  
Bilayer $^{3}He$,
is in essence, a localized layer of Heisenberg spins, coupled vertically
via a Kondo coupling to an upper layer of mobile fermions.
One of the challenges this poses, is to connect the phase diagram of the two dimensional quantum
antiferromagnet, with the two dimensional Kondo lattice (Fig. \ref{fig7} )

There are two ways for zero point quantum fluctuations to melt 
the  magnetic order in a quantum antiferromagnet:
\begin{itemize}

\item by increasing the zero-point fluctuations of the
spin (by reducing the size $S$ of the spin or increasing the magnetic
frustration), or

\item by coupling
the spin fluid via a Kondo coupling to a two-dimensional fluid of
fermions. 

\end{itemize}
These two processes define two orthogonal axes of a phase diagram, as
shown in Fig. \ref{fig7}.
Along the ``frustration (x-) axis'' that loosely defines
insulating antiferromagnets, there is a quantum
phase transition from a spin liquid to an ordered two dimensional antiferromagnet. This
quantum phase transition has been extensively studied, and is
the source of inspiration for ``deconfined
quantum criticality'' advanced by Senthil et al.  as a theory 
for critical behavior which is dominated by the
formation of a fluid of deconfined spinons\cite{senthilfisher}. 
Along the Kondo (y-) axis, there is the quantum phase transition from antiferromagnet to heavy
electron Fermi liquid. 
A working mean-field theory would show us 
how these two phase transitions are connected.  For highly frustrated, low 
spin lattices, we will also
have to understand how a gapped spin liquid  can make a transition to
the heavy Fermi liquid, and whether, as seems the case in both $MnSi$\cite{mnsi}
and $UCoAl$\cite{ucoal}, there is an intermediate non-Fermi liquid ground-state
separating the antiferromagnet from the fully developed Fermi liquid.

There are some interesting technical practical problems to be solved 
to actually realize this phase diagram in a controlled calculation.
One important aspect to the
problem, is finding a way to control the fluctuations to form
a mean-field theory. In classical criticality, this kind of control
was obtained by generalizing the three component magnetization
to an $N$ component order parameter, the
so called ``spherical model'', where the $O (3)$ rotation group  is
replaced by the $O (N)$ rotation group\cite{kac,wilsonspherical}.  In heavy fermion physics, we
have yet to find the appropriate large $N$ expansion to described the
phases on both sides of the quantum critical point. For a long time,
our field has followed the example of the particle physics, in trying
to use the $SU (N)$ group for this purpose.   
The problem with $SU (N)$
group, is that singlets  are baryons, and they require complexes of
$N$ particles, 
\begin{equation}
\vert \hbox{singlet SU (N)}\rangle =f^{\dagger }
_{1}f^{\dagger }_{2}\dots  f^{\dagger }_N\vert  0 \rangle 
\end{equation}
But in condensed matter physics, the formation of
singlets between pairs of particles - electrons in superconductors,
spins in antiferromagnets, plays a very important role.
An important group for our purposes, may well be the group $SP
(2N)$, which preserves the special relationship between up and down
electrons\cite{readsachdev}. In SP (2N), the spin labels on the states are given by
$(\uparrow, a)$ and $(\downarrow,a )$, where $a$ runs from $1$ to
$N$, and a pair operator still defines a singlet
\begin{equation}
\vert \hbox{singlet SP (2N)}\rangle = \sum_{a=1,N}f^{\dagger}
_{\uparrow,a}f^{\dagger }_{\downarrow,a }\vert  0 \rangle 
\end{equation}
One of the nice properties of this group, is that it preserves
a local $SU (2)$ particle-hole gauge
symmetry\cite{affleck,colemanandrei}, 
yet surprisingly, this symmetry has has not yet 
been applied to the Kondo impurity, or Kondo lattice problems. 

Another important technical problem, is
how to tune $S$, as one does in insulating antiferromagnets, while
maintaining a perfectly screened Kondo lattice.  This part of the
problem was fortunately solved some years ago by Parcollet and
Georges, who pointed out that that one could consider a family
of Kondo lattice models of different spin $S$, in which each spin is
coupled to $2S$ screening channels\cite{parcolletgeorges}.   
The ``mean-field theory''
that appears in this kind of Eliashberg or dynamical mean field theory.
Recent results of Jerome Rech at Rutgers, and my
collaborators Gergely Zarand and Olivier Parcollet indicate
\cite{rech,indranil}
 that it
may be possible to exactly compute the mean-field phase diagram in
the large $N$ limit using this scheme, using 
Schwinger bosons. Schwinger bosons are an effective method to
describe low dimensional antiferromagnetism, and our most recent work
confirms that they can also describe the fully screened Fermi liquid
physics of the one and two impurity Kondo model.   
One of the exciting
aspects of this kind of approach, is that the initial mathematical
formulation involves a gauge theory of electrons interacting with spinons,
which carry the local moment spin, and charged spinless fermions, or
``holons''  which mediate the Kondo
interaction.
Remarkably, when one solves the equations for the fully screened Kondo
model, one finds that 
spinon and holons develop a gap in the Fermi liquid. 
Antiferromagnetism corresponds to the condensation of spinons, and 
when this happens, the gap for both holons and spinons must collapse
suggesting a kind of critical
behavior featuring a co-existence of critical spin and charge
fluctuations (see Fig.\ref{fig7}.).
Taken seriously, this appears to predict that a heavy electron Fermi liquid
near to a quantum critical point may have new, low lying spinon and
holon excitations. 
We hope to report on these results in future meetings. 

\section{Postscript.}\label{}

I came to my first conference as a new Ph.D., twenty one years
ago, in Cologne 1984, just as the ideas of heavy electron bands, 
and heavy electron
superconductivity were becoming  accepted phenomena.  I was excited,
had the sense that realistically, we'd probably had our fair quota of
big discoveries.
Had anyone tried to convince me that d-wave superconductivity at  $135K$
superconductivity, quantum dots, routine experiments at $10^{-6}K$ in
things called atom traps, and a completely new class of phase transition,
would be part of the near future, 
I would certainly have dismissed it as wonderful science fiction!
Yet all of this happened in the past twenty years. So perhaps you
today - as a student are wondering, if now, finally, the big discoveries are past?

No! Absolutely not! 
Today, the reasons for optimism about research in strongly correlated
electrons are at least as bright as twenty years ago, and if the past
is any lesson for the
future, major discoveries - both experimental and theoretical, will come
that will continue to outstrip our expectation.
I think think as experimentalists and theorists, especially the
students in the audience - we can not be too ambitious
with our research at this time. It is still an exciting time, and I wish you
good spirits and good luck as you return to your home institutions.

\vspace{0.3cm}

\section*{Acknowledgement}
This research was supported by the
National Science Foundation, USA,  grant  DMR-0312495. I am indebted
to John Saunders, for discussions about his bilayer $^{3}He$ data, and
to Neil Harrison and Jaime Moreno for providing figure 2. at very short notice.
Thanks to Stephen Julian for drawing the possible non-Fermi liquid
phase of $UCoAl$ to my attention.


\begin{thebibliography}{99}

\bibitem{rochester}{\it Valence Instabilities and Narrow Band
Phenomena}, edited by R. Parks, (Plenum 1977).


\bibitem{nexus} K. H. Kim, N. Harrison, M. Jaime, G. S. Boebinger and 
 J. A. Mydosh, Phys. Rev. Lett. {\bf 91}, 256401 (2003)

\bibitem{harrison}N. Harrison, A. V. Silhanek, C. D. Batista,
M. Jaime, A. Lacerda, H. Amitsuka, J. A. Mydosh, {\bf Th-FHO-5}, p. 104.

\bibitem{satovienna}N. K. Sato, S. Uemura, G. Motoyama and
T. Nishioka, {\bf Th-FHO-10}, p 170. 

\bibitem{bernal_vienna}O. Bernal, M. E. Moroz, H. Murukawa,
A. P. Reyes, P. L. Kuhns, D. E. MacLaughlin, H. G. Lukefahr,
J. A. Mydosh, T. J. Gortenmulder and H. Amitsuka, {\bf Th-FHO-9},p 170.

\bibitem{zhitormsky}M. E. Zhitormsky and  V. P. Mineev, {\bf
Th-FHO-6}, p. 105.

\bibitem{bauer}E. Bauer et al, Physical Review Letters, {\bf 92}, 
027003/1-4, (2004).

\bibitem{cept3si1}Danierl Agterburg, Raminder P. Kaur, Manfred
Sigrist, Paolo Frigeri and Akihisa Koga, {\bf Th-SCO-6}, 98.

\bibitem{cept3si2} 
Mamoru Yogi, Hidekazu Mukuda, Yoshio Kitaoka, Shi Hashimoto, Takashi Yasuda, Rikio Settai, Tatsuma D. Matsuda, Yoshimori
Haga, Yoshichika Onuki, Peter Rogl and Ernst Bauer, {\bf Th-SCO-7
}, p 99. 

\bibitem{cept3si3}P. A. Frigeri, M. Sigrist and D. Agterburg, {\bf
Fr-NSC-35}, p 272.

\bibitem{saunders2}M. Neumann, Andrew Casey, Jan Ny\' eki, Brian Cowan
and John Saunders, {\sl ``
$^{3}He$ Films as Model Strongly Correlated
Fermion Systems: observation of a magnetic quantum critical point"},
{\bf Fr-HF-5}, (191).

\bibitem{saunders1}A. Casey, H. Patel, J. Ny\' eki, B. P. Cowan, and J. Saunders,
Phys. Rev. Lett. {\bf 90}, 115301/1-4 (2003).
%^{3}He single layer

\bibitem{grigera}R. S. Perry, A. J. Schofield, A.J., M.  Chiao, S. R.
Julian, G. G.  Lonzarich, S. Ikeda, Y. Maeno, A. J.  Millis, and
A. P. Mackenzie,  Science {\bf 294}, 329 (2001).

\bibitem{hidden_order} V. Tripathi, P. Chandra and  P. Coleman, 
J. Phys.: Condens. Matter {\bf 17}, 5285 (2005).


\bibitem{bernal_old}O.O. Bernal, C.  Rodrigues, A.  Martinez, H. G. Lukefahr,
D. E. MacLaughlin, A. A. Menovsky, and J. A.  Mydosh, Phys. 
Rev.  Lett. {\bf 87}, 153, (2001).

\bibitem{cept3si2nmr}M. Yogi, Y. Kitaoka, S. Hashimoto, T. Yasuda, R. Settai, T. D. Matsuda, Y. Haga, Y. Onuki, P. Rogl, and E. Bauer
Phys. Rev. Lett. {\bf 93}, 027003 (2004).

\bibitem{affleck}I. Affleck, Z. Zou, T. Hsu and P. W. Anderson , {\sl  `` Local SU(2)
symmetry of the Heisenberg model''}, 
Phys. Rev. B {\bf 38} 754-759 (1988).


\bibitem{colemanandrei}P. Coleman
and N. Andrei, J. Phys. Cond. Matt {\bf C 1.}, 4057-4080 (1989).

\bibitem{landau}L. D. Landau, {\sl ``Theory of Fermi Liquids''},
 Zh. Eksperim. i Teor. Fiz. {\bf 30}, 1058
(1956); [English transl:
Soviet Phys. [JETP 3, 920 (1956)].


\bibitem{greywall}D. S. Greywall and P. A. Busch, Phys. Rev. Lett. 
{\bf 62}, 1868-1871 (1989).


\bibitem{wilson}K. G. Wilson, Rev. Mod. Phys. {\bf 47}, 773, (1976).


\bibitem{white}Steven R. White, Phys. Rev. Lett. {\bf 69}, 2863 (1992).

\bibitem{hotta}Chisa Hotta and Naokazu Shibata, {\bf Fr-HF-36}, p 209.

\bibitem{doniach}S. Doniach, {\it Valence Instabilities and Narrow Band
Phenomena}, edited by R. Parks, 34, (Plenum 1977); S. Doniach, {\it Physica }{\bf B91}, 231 (1977).


\bibitem{zerec}Iviva Zerec, Burkhard Schmidt, Peter Thalmeier and
Peter Fulde, Fr-HF-32, p 207. 

\bibitem{dmft}A. Georges, G. Kotliar, W. Krauth,
and MJ Rozenberg, Rev. Mod. Phys. {\bf 68}, 13 (1996). 

\bibitem{koller}Winfried Koller, Alex Hewson, David Edwards and
Dietrich Meyer {\bf Fr-HF-29}, p205.

\bibitem{hanketal}Werner Hanke, Markus Aichorn, Christopher Dahnken,
Enrigo Arrigoni and Michael Potthoff, {\bf W2-SCE1-2},p 11.


\bibitem{vollhardt}D. Vollhardt, WE-SCE1-1, p10. 


\bibitem{rosch}Achim Rosch, {\bf  W2 QC-3}, p5. 

\bibitem{roschsi} Lijun Zhu, Markus Garst, Achim Rosch and  Qimiao Si,
Phys. Rev. Lett. {\bf 91}, 066404 (2003).
 

\bibitem{kadowakiwoods} K. Kadowaki and S. B. Woods, Solid State
Commun. {\bf 58}, 507 (1986).


\bibitem{tsujii}N. Tsujii, H. Kontani and K. Yoshimura {\bf Fr-HF-54}, p218.



\bibitem{tsujii2}N. Tsujii, H. Kontani and K. Yoshimura, Phys. Rev. Lett. {\bf  94}, 057201 (2005). 

\bibitem{gegenwart1}Frankziska Weickert, Julia Ferstl, Teodara Radu,
Philipp Gegenwart, Christoph Geibel and Frank Steglich, {\bf
We-QC-11}, p 14.

\bibitem{gegenwart2}P. Gegenwart, J. Custers, Y. Tokiwa, C. Geibel, and F. Steglich
Phys. Rev. Lett. {\bf 94}, 076402 (2005).

\bibitem{slebarski1}
Andrzej Slebarski and Jozef Spalek, {\bf
We-NFL-7}, p 41.

\bibitem{slebarski2}A. Slebarski and J. Spalek, Phys. Rev. Lett. {\bf 95}, 046402 (2005).


\bibitem{fflo1}H. A. Radovan, N. A. Fortune, T. P. Murphy,
S. T. Hannahs, E. C. Palm, S. W. Tozer and D. Hall, 
Nature {\bf 425}, 51-55 (2003).

\bibitem{fflo2}A. Bianchi, R. Movshovich, C. Capan, A. Lacerda,
P. G. Pagliuso and J. L. Sarrao, Phys. Rev. Lett {\bf 91 },187004/1-4,
(2003).


\bibitem{radovan}H. A. Radovan, {\bf  Th-SCO-3}, p 97.


\bibitem{fflo3}K. Kumagai, H. Kakuyanagi, M. Saito, K. Kumagai,
S. Takashina, M. Nohara, H. Takagi, Y. Matsuda, {\bf Th-SCO-4}, p97.

\bibitem{fflobdg}See e.g.
S. Matsuo, Y. Nagato and K.  Nagai K., 
{\sl ``Phase diagram of the Fulde-Ferrell-Larkin-Ovchinnikov state in a three-dimensional superconductor''}
Journal of the Physical Society of Japan, {\bf 67}, 280,  (1998). 


\bibitem{qc1}Piers Coleman and Andrew J. Schofield, Nature {\bf 433}, 226-229 (2005).

\bibitem{qc2}R. B. Laughlin, G. G.  Lonzarich, 
P. Monthoux and  D.  Pines, {\sl  ``The quantum criticality conundrum''} {\it Advances in Physics}, 
{\bf 50}, 361-5 (2001).


\bibitem{review}P. Coleman, P., C. P\'epin, C., Q. Si and R.  Ramazashvili,.
{\em {J.\ Phys.: Condens.\ Matter}}{ \bf 13}, R723--R738 (2001).

\bibitem{rowley}Stephen E.  Rowley, S.S. Saxena and G. G. Lonzarich,
post-deadline poster. 

\bibitem{miyake}Kazumasha Miyake, {\bf  Th-P-1}, p 95.

\bibitem{cecu2gesi} A. T. Holmes , D. Jaccard  and K. Miyake, 
{\sl `` Signatures of valence fluctuations in 
$CeCu_{2}Si_{2} $ under high pressure''}, Phys. Rev. B {\bf  69}, 024508 (2004).

\bibitem{fisher}Y. Matsushita, H. Bluhm, T. H. Geballe, and I. R. Fisher, {\sl ``Evidence for Charge Kondo Effect in Superconducting
Tl-Doped PbTe''},Phys. Rev. Lett. {\bf 94}, 157002/1-4 (2005).

\bibitem{dzero}M. Dzero and J. Schmalian,{\sl ``Superconductivity in Charge Kondo Systems''},
Phys. Rev. Lett. {\bf 94}, 157003/1-4 (2005)


\bibitem{qimiao}Qimiao Si, {\bf We-P-1}, p3. 

\bibitem{hertz} J. A. Hertz, Phys. Rev. B {\bf 14}, 1165 (1976).

\bibitem{moriya}T. Moriya and J. Kawabata, J. Phys. Soc. Japan {\bf
34}, 639 (1973); J. Phys. Soc. Japan {\bf 35},669 (1973).

\bibitem{si} Q. Si, S. Rabello, K. Ingersent and J. L. Smith, Nature

\bibitem{pepin05}C. P\' epin, Phys. Rev. Lett. {\bf 94}, 066402 (2005)


\bibitem{senthilfisher}T. Senthil, Ashvin Vishwanath, Leon Balents, Subir
Sachdev, M. P. A. Fisher, 
Science 303, 1490-4 (2004).

\bibitem{mnsi}N. Doiron-Leyraud, I. R. Walker, L.
Taillefer. M. J. Steiner, 
S. R. Julian, G. G.  Lonzarich, Nature {\bf 425}, 595 (2003).

\bibitem{ucoal}A. J. Bograd, J. S. Alwood, D. J. Mixon, J. S. Kim and
G. R. Stewart, www.phys.ufl.edu/REU/2002/reports/bograd.pdf unpublished.


\bibitem{kac} 
T. H. Berlin and M. Kac, {\sl ``The Spherical Model of a Ferromagnet''}
Phys. Rev. {\bf 86}, 821-835 (1952).

\bibitem{wilsonspherical}K. G. Wilson, Phys. Rev. D 7, 2911 (1973).

\bibitem{readsachdev} Subir Sachdev and N. Read, 
{\sl ``Large N expansion for frustrated and doped quantum antiferromagnets''},
International Journal of Modern Physics B {\bf 5}, 219 (1991).

\bibitem{parcolletgeorges}O. Parcollet and A. Georges, PRL {\bf 79}, 4665-8
(1997).

\bibitem{rech}J. Rech, P. Coleman, O. Parcollet and
G. Zarand, Phys. Rev. Lett. in press (2005).

\bibitem{indranil}P. Coleman, I. Paul and J. Rech, Phys. Rev. {\bf B 72}, 094430 (2005).

\end{thebibliography}
\end{document}